\documentclass[a4paper, 12pt]{article}

\newcounter{qcounter}
\usepackage{graphicx}

\usepackage{subfigure}

\usepackage[round]{natbib}

\usepackage{acronym}
\usepackage{url}
\acrodef{PCG}{Projected Conjugate Gradient} 
\acrodef{QP}{quadratic programming}
\acrodef{RBF}{Radial-Basis Function}
\acrodef{ABM}{Agent-Based Modelling}
\acrodef{AI}{Artificial Intelligence}
\acrodef{DAI}{Distributed Artificial Intelligence}
\acrodef{API}{Application Programming Interface}
\acrodef{ARF}{p14ARF human tumor-suppressor gene}
\acrodef{B2B}{business-to-business}
\acrodef{BDP}{Biological Design Pattern}
\acrodef{BGS}{Best Guess Solution}
\acrodef{BIC}{Biologically-Inspired Computing}
\acrodef{BML}{Business Modelling Language}
\acrodef{BPEL}{Business Process Execution Language}
\acrodef{BPMN}{Business Process Modelling Notation}
\acrodef{CAS}{Complex Adaptive Systems}
\acrodef{COBOL}{COmmon Business-Oriented Language}
\acrodef{DBE}{Digital Business Ecosystem}
\acrodef{DE}{Digital Ecosystem}
\acrodef{DEC}{distributed evolutionary computing}
\acrodef{DGA}{Distributed genetic algorithms}
\acrodef{DIS}{Distributed Intelligence System}
\acrodef{DNA}{Deoxyribose Nucleic Acid}
\acrodef{DOP}{DBE Open Protocol}
\acrodef{DSS}{Distributed Storage System}
\acrodef{EAP}{Evolving Agent Population}
\acrodef{ebXML}{e-business eXtensible Markup Language}
\acrodef{EC}{Evolutionary Computing}
\acrodef{ECJ}{Evolutionary Computing in Java}
\acrodef{EE}{Evolutionary Environment}
\acrodef{EFL}{Evolutionary Framework for Language}
\acrodef{FLE}{Framework for Language Ecosystems}
\acrodef{EOA}{Ecosystem-Oriented Architecture}
\acrodef{ESS}{evolutionary stable strategy}
\acrodef{EvE}{Evolutionary Environment}
\acrodef{ExE}{Execution Environment}
\acrodef{FCB}{Framework for Computational Biomimicry}
\acrodef{FFF}{Fitness Function Framework}
\acrodef{FL}{Fitness Landscape}
\acrodef{HWU}{Heriot-Watt University}
\acrodef{ICL}{Imperial College London}
\acrodef{ICT}{Information and Communications Technology}
\acrodef{INTEL}{Intel Ireland}
\acrodef{IPA}{International Phonetic Alphabet}
\acrodef{ISUFI}{Istituto Superiore Universitario di Formazione Interdisciplinare}
\acrodef{JDJ}{Java Developer's Journal}
\acrodef{KC}{Kolmogorov-Chaitin}
\acrodef{LAN}{local area network}
\acrodef{LSE}{London School of Economics and Political Science}
\acrodef{MAS}{Multi-Agent System}
\acrodef{MDL}{Minimum Description Length}
\acrodef{MDM2}{murine double minute 2}
\acrodef{MFT}{Mean Field Theory}
\acrodef{MoAS}{Mobile Agent System}
\acrodef{MOF}{Meta Object Facility}
\acrodef{MUH}{migration and usage history}
\acrodef{NIC}{Nature Inspired Computing}
\acrodef{NN}{Neural Network}
\acrodef{NoE}{Network of Excellence}
\acrodef{OMG}{Open Mac Grid}
\acrodef{OPAALS}{Open Philosophies for Associative Autopoietic Digital Ecosystems}
\acrodef{P2P}{peer-to-peer}
\acrodef{P53}{protein 53}
\acrodef{PDA}{Personal Digital Assistant}
\acrodef{QoS}{quality of service}
\acrodef{REST}{REpresentational State Transfer}
\acrodef{RNA}{Deoxyribose Nucleic Acid}
\acrodef{SAE}{Software Agent Ecosystem}
\acrodef{SBML}{Systems Biology Modelling Language}
\acrodef{SBVR}{Semantics of Business Vocabulary and Business Rules}
\acrodef{SDL}{Service Description Language}
\acrodef{SF}{Service Factory}
\acrodef{SIM}{Social Interaction Mechanism}
\acrodef{SM}{Service Manifest}
\acrodef{SME}{Small and Medium sized Enterprise}
\acrodef{SML}{Service Modelling Language}
\acrodef{SMO}{Sequential Minimal Optimisation}
\acrodef{SOA}{Service-Oriented Architecture}
\acrodef{SOAP}{Simple Object Access Protocol}
\acrodef{SOC}{Self-Organised Criticality}
\acrodef{SOLUTA}{SOLUTA.NET}
\acrodef{SOM}{Self-Organising Map}
\acrodef{SSL}{Semantic Service Language}
\acrodef{STU}{Salzburg Technical University}
\acrodef{SUN}{Sun Microsystems}
\acrodef{SVM}{Support Vector Machine}
\acrodef{TM}{Turing Machine}
\acrodef{UBHAM}{University of Birmingham}
\acrodef{UDDI}{Universal Description Discovery and Integration}
\acrodef{UML}{Unified Modelling Language}
\acrodef{URI}{Uniform Resource Identifier}
\acrodef{UTM}{Universal Turing Machine}
\acrodef{VLP}{variable length population}
\acrodef{VLS}{variable length sequences}
\acrodef{vls}{variable length sequence}
\acrodef{WP}{Work-Package}
\acrodef{WSDL}{Web Services Definition Language}
\acrodef{XMI}{XML Metadata Interchange}
\acrodef{XML}{eXtensible Markup Language}
\acrodef{MD5}{Message-Digest algorithm 5}
\acrodef{GA}{genetic algorithm}
\acrodef{GP}{genetic programming}
\acrodef{MASON}{Multi-Agent Simulator Of Neighbourhoods}
\acrodef{Repast}{Recursive Porous Agent Simulation Toolkit}
\acrodef{JCLEC}{Java Computing Library for Evolutionary Computing}
\acrodef{OWL-S}{Web Ontology Language - Service}
\acrodef{EGT}{Evolutionary Game Theory}
\acrodef{RBF}{Radial Basis Functions}
\acrodef{SWS}{Semantic Web Services}
\acrodef{HDD}{Hard Disk Drive}
\acrodef{SSD}{Solid-State Drive}
\acrodef{OKS}{Open Knowledge Space}
\acrodef{CAES}{Complex Adaptive EcoSystem}
\acrodef{SaaS}{Software-as-a-Service}
\acrodef{PaaS}{Platform-as-a-Service}
\acrodef{IaaS}{Infrastructure-as-a-Service}
\acrodef{C3}{Community Cloud Computing}
\acrodef{SSO}{single sign-on}

\usepackage{color}
\definecolor{gray}{rgb}{0.5,0.5,0.5}

\usepackage[paper=a4paper, margin=2.5cm]{geometry}

\usepackage[
pdftitle={Digital Identity in The Absence of Authorities: A New Socio-Technical Approach},
pdfauthor={Gerard Briscoe},
colorlinks=true, linkcolor=black, citecolor=black, filecolor=black, urlcolor=black,
hyperfootnotes=false]{hyperref} 

\begin{document}

\title{Digital Identity in The Absence of Authorities:\\A New Socio-Technical Approach}

\author{Mark McLaughlin\thanks{Telecommunications Software \& Systems Group, Waterford Institute of Technology, Ireland, \{mmclaughlin, pmalone\}@tssg.org} \and Gerard Briscoe\thanks{Department of Media and Communications, London School of Economics and Political Science, United Kingdom, g.briscoe@lse.ac.uk} \and Paul Malone$^*$}

\date{}

\maketitle

\begin{abstract}
On the Internet large service providers tend to control the digital identities of users. These defacto identity authorities wield significant power over users, compelling them to comply with non-negotiable terms, before access to services is granted. In doing so, users expose themselves to privacy risks, manipulation and exploitation via direct marketing. Against this backdrop, the emerging areas of Digital Ecosystems and user-centric identity emphasise decentralised environments with independent self-determining entities that control their own data and identity. We show that recent advances in user-centric identity, federated identity and trust have prepared the ground for decentralised identity provisioning. We show how social trust, rather than blind deference to authorities, can provide a basis for identity, where risks can be weighed and compared rather than merely accepted. Fundamentally, we are considering the move from authority-centric centralised identity provisioning to user-centric distributed identity provisioning. Finally, we highlight the potential impacts of distributed identity provisioning in the Information Society and give a brief roadmap for its general implementation and adoption.\\

\end{abstract}

\section{Introduction}
\label{intro}
This paper is concerned with digital identity\footnote{Henceforth we abbreviate the term to `identity'.} in decentralised environments, where identity authorities either do not exist or play a limited role. We have two decentralised environments in mind: i) the web, where users access services on the Internet via a web browser, and ii) Digital Ecosystem platforms where users use enhanced clients to access web services via arbitrary service access protocols. Usually, our analysis applies equally to both cases; where this is otherwise, a distinction is made.

The ethos of Digital Ecosystems\footnote{http://www.digital-ecosystems.org/} (DEs) favours open, distributed service platforms as an alternative to the `keystone' model, in which entities cluster around systems that are owned and maintained by a small number of authoritative entities. DEs are composed of distributed, interconnected groups of equal entities, in contrast to the keystone model that fosters an undesirable dependence of `ordinary' entities on `authoritative' entities. DEs are effective when they foster broad, diverse organisations of entities that are free to compete and collaborate based on dynamic social factors. Identity in DEs must be similarly distributed and decentralised, and founded on social relationships within the ecosystem. The virtual organisation of identities in an ecosystem can be described as emergent, decentralised, informal, and though based on local relationships is potentially global in extent.

Traditionally, identity has been concerned almost solely with the use of a username-password pair to authenticate a user for access to a service. A small set of data items stored by the service provider (SP) determines the user's access rights and other information pertinent to service use. Users, typically, have one identity per service, and the SP provides the identity, via an internal identity provider (IdP). The IdP is responsible for retaining information pertaining to this identity, providing authentication and authorisation, and generally speaking presiding over the entire life-cycle of the identity. For each service, there is a prescribed IdP that users must deal with. Where services become very popular, and where services proliferate on the same network, or service platform (e.g. Google, Yahoo!), a common IdP is invariably used to manage identities for all services. These IdPs are identity authorities for those environments and users are compelled to accept their terms of service and to trust them with supplied personal information.

Recent developments in identity, driven by technical, social and business concerns have begun to change this \emph{landscape} significantly. User-centric identity developments have led to the logical and functional separation of SP and IdP, which allows the user to choose the IdP that provides their identity to an SP. Federated Identity is concerned with linking identity domains intra- and inter-organisation such that identities on one system can be used to access services on another. Considerable research has been conducted on privacy preserving identity management, which highlights the sensitivity of identity to interference by identity authorities and data controllers. Other research has recognised that traditional identity management systems have tended to give rise to unequal power relations that place the SP/IdP at an advantage to the user. These developments have laid the foundations for a fundamentally different kind of identity management, that is not backed and controlled by authorities but is backed by trust and controlled by the user. 

In contexts where reliance on identity authorities is unnecessary or undesirable, we propose the move from authority-centric centralised identity provisioning \citep{miyata_surveyidentity_2006} to user-centric \citep{maler_venn_2008} distributed identity provisioning \citep{markmcl_2009}, in which identity is provided based on trust, ultimately derived through social networks.
\section{Literature and technology review}
\label{sec:1}

\subsection{Key concepts}

\hspace{0.5cm} \emph{Digital identity} \citep{glasser_identity_2008,cameron,jsang_user_2005,pfitzmann_anonymity_2005} is concerned with how people are identified on computer systems and the internet\footnote{Identity management, comprising the protocols and use of credentials to establish identity, is a major area in its own right and we do not discuss it directly here.}. Partial identities are `that which represents a person in a particular context in the online world' \citep{hansen_privacy-enhancing_2004,glasser_identity_2008}, which is the type of identity we will use to represent users in practice. Identity is traditionally responsible for determining access rights to sensitive resources for users. 
With the advent of Web 2.0, web services have appeared that foster the development of rich online identities comprising personal attributes, preferences and behaviour, in addition to access related metadata. For example, social networking sites such as Facebook encourage users to create detailed user profiles and to replicate, and build on, real world social networks on its site. 
This represents a shift from a purely technical and security oriented concept of identity to one that is socio-technical and enabling of social services.

\textit{Federated identity} is concerned with federating previously separate identity domains, across large organisations and the enterprise, such that users in one domain can consume services in another. ``Federated identity infrastructure enables cross-boundary \ac{SSO}, dynamic user provisioning and identity attribute sharing. By providing for identity portability, identity federation affords end-users with increased simplicity and control over the movement of personal identity information while simultaneously enabling companies to extend their security perimeter to trusted partners." \citep{sourceid_saml_2009}.

\textit{User-centric identity} is a philosophy, and set of supporting standards and technologies, for empowering the user by giving them control over their identities \citep{maler_venn_2008}. The major philosophical innovation is in forming a distinction between SP and IdP in online applications, where the two have always been seamlessly integrated, and providing technologies for allowing those SPs (e.g. Facebook, Twitter) to use an IdP of the user's choice (rather than its own) to authenticate the user for the service. This move has given users the opportunity to manage consent for personal data disclosure, manage their own credentials, and perform authentication, or \ac{SSO} (where the IdP will authenticate the user to a range of services), independently of SPs\footnote{Though only if SPs support a user-centric sign-on protocol.} (e.g. OpenID\footnote{http://openid.net/}). 

\textit{Trust} has long been a topic of study in psychology, sociology, philosophy and economics; but in the nineties it has also found application in e-commerce, particularly in online markets such as eBay \citep{sabater_reviewcomputational_2005}. Trust can be described as ``a directional relationship between two parties that can be called \textit{trustor} and \textit{trustee}," \citep{jsang_trust_2007} where a trustor is said to trust, or not to trust, a trustee, in a particular context. Trust can be used as a form of `soft security' \citep{jsang_trust_2007} or, by reflecting the real world social relations, as an enabler of ``trade, competition, collaboration and so on" \citep{sabater_reviewcomputational_2005}. There are numerous models for computing trust and reputation\footnote{``The overall quality or character [of some trustee] as seen or judged by people in general." \citep{jsang_trust_2007}} \citep{sabater_reviewcomputational_2005,jsang_trust_2007} on various systems and networks, including decentralised P2P networks \citep{marti_taxonomy_2006}.

\textit{Trust transitivity} describes how trust propagates on social networks, and is predicated on the principle that if A trusts B and B trusts C, A indirectly trusts C under certain conditions \citep{huang_ontology_2006}. Trust transitivity can be useful where one party may not have a direct trust relationship with another party. For example, even though A may not know a `good' mechanic, A may accept a recommendation, or referral, from B recommending a good mechanic C. 'Referral' trust, as a special case of trust in `belief' \citep{huang_ontology_2006}, is transitive \citep{jsang_trust_2007}, meaning that the beliefs of parties are trusted, and can be passed on to others in a chain of trust. These trusted beliefs feed into decision making in the real world, and recently in the online world as well.

\textit{Trust networks} can be built from transitive trust chains \citep{jsang_trust_2006}. Directed `arcs' in the network, representing trust relationships between the trustor `node' and the trustee node, can be `coloured' to indicate the trust context. Contextual trust chains can be traced by following same coloured arcs of `belief' trust from a `source' node, the trustor, to a `sink' node, the trustee\footnote{The final arc in the chain must actually be a `performance' \citep{huang_ontology_2006} or `functional' \citep{jsang_trust_2007} trust arc.}. 

\textit{Digital Ecosystems} can be described as ``distributed adaptive open socio-technical system, with properties of self-organisation, scalability and sustainability, inspired by natural ecosystems" \citep{bionetics}. The term has been used in other contexts \citep{thesis}, however, we refer here to the body of research initiated under the heading of Digital Business Ecosystems (DBE), that is intended to promote the pervasiveness of \ac{ICT} in \acp{SME} and move organisations towards a ``more, fluid amorphous and, often, transitory structures based on alliances, partnerships and collaboration" using biologically inspired metaphors \citep{nachira_towardsnetwork_2002}. This contrasts with the `keystone' model, where smaller supplier firms cluster around one large firm, which is a successful business structure only when the major actor is economically `healthy'. DEs support DBEs as a particular case, i.e. business \citep{bionetics}. The OPAALS\footnote{\acl{OPAALS} \citep{opaals}} approach to creating a DE infrastructure is to build a decentralised service platform from P2P and other distributed technologies, informed by social theory and biological metaphors \citep{dini2008bid}. Several strands of research from DEs are relevant to our efforts, including the theory of power structures in identity systems, privacy preserving identity management and user-centric identity emphasise the imbalance of power between large SPs with in-house IdPs, such as Google on the web, and the government and enterprises on other networks; and the users that consume their services.

\subsection{Decentralised identity}

The following principles of DEs influence our approach to decentralised identity provisioning, namely, ``no single point of failure or control," ``should not be dependent upon any single instance or actor," ``equal opportunity of access for all," ``scalability and robustness," ``ability to evolve, differentiate, and self-organise constantly," and ``local autonomy" \citep{nachira_digital_2007}. 

The `carrot', for users, of a more equal and more sustainable ecosystem of services is re-enforced by the `stick' of an emerging surveillance society \citep{pounder_nine_2008}, where undesirable power relations are developing between controllers of personal information and users, that make users suspicious of authorities. The associated risks to the user include erosion of privacy, the influence and manipulation of persons or populations, and user-profiling activities \citep{halperin_roadmap_2008}. A sociological survey and analysis of privacy issues and `power' relations between user and SP/IdP in collaborative workspaces and social networks is given by \citet{pekrek_comparison_2009}, who conclude that ``social norms are currently the only forces effectively delimiting the unabridged use of personal information," and cites the use of open source P2P networks as an improvement over centralised servers for data privacy. Furthermore, \citet{krasnova_privacy_2008} shows that concerns about privacy on social networks can threaten their long term viability. Even from a sheer practical point of view, the risk of data breaches, through theft, accidental exposure, viruses and hacking, or insufficient access control and security put user data at risk when data controllers hold large amounts of data \citep{verizon_data_2008}\footnote{``...organizations that process or store large quantities of data valued by the criminal community; they are the quintessential Targets of Choice."}.

Privacy enhancing identity management comprises an important stream of European research, starting with PRIME\footnote{https://www.prime-project.eu/}, whose guiding principle was to ``put individuals in control of their personal data" \citep{leenes_prime_2008}, and is concerned with issues such as minimising data disclosure and negotiating privacy policies with SPs. Similar concerns are expressed by Cameron's `Laws of Identity' \citep{cameron}. This research echoes and reinforces from a theoretical perspective what user-centric identity specifications and technologies are accomplishing at a practical level.

\subsection{Technology}
The two main federation specifications are SAML v2.0\footnote{http://docs.oasis-open.org/security/saml/v2.0/saml-core-2.0-os.pdf} and WS-Federation\footnote{http://www.ibm.com/developerworks/library/specification/ws-fed/}, both of which are OASIS\footnote{http://www.oasis-open.org/} standards. SAML v2.0 is the older specification and enjoys the most widespread adoption. Shibboleth\footnote{http://shibboleth.internet2.edu/} has become the defacto SAML compliant, federated identity solution and is open source.

A model for identity in DEs was outlined by \citet{markmcl_2009}, that is capable of building arbitrary identity operations, which are protocols that conduct identity tasks such as sign-on. The model is backed by trust rather than by a central authority, is agnostic of particular platform implementations, and can operate in purely decentralised environments such as P2P. \textit{IdentityFlow}\footnote{http://sourceforge.net/projects/identityflow/} is an open source implementation of this identity model.
\section{Trust requirements for identity}
\label{sec:2}
Identity is always accepted by third parties on the basis of trust, however until recently \citep{kylau_trust_2009,jsang_trust_2005} the trust requirements of identity have not been analysed explicitly. It was sufficient for an SP, combined with an authoritative identity provider, to control the identity of users and to compel them to authenticate each time it consumed the service. The user was expected to either blindly trust the SP/IdP or not to participate in the service\footnote{Although, it is true that the use of public key infrastructure (PKI) certificates can give users some assurances by linking the SP to a real world entity.}. In effect, the SP reduced its risk to almost zero, while the user's risk was 
hardly considered. With the advent of federated and user-centric identity, and the possibilities of decentralised identity, a more formal analysis is required. By understanding how trust is required in these scenarios, we can give requirements for trust in decentralised identity.

Identity is asserted either by the subject, i.e. the user, or by some representative of the subject, such as the user's IdP; this identity assertion, or claim, is then verified by a relying party, i.e. an SP, or by a third party on the relying party's behalf, such as the relying party's IdP. These two statements on identity assertion and verification will hold true in every scenario. What will vary is the exact configuration of actors participating in an \textit{identity task}. We define an identity task as the task of generating an identity assertion and making it acceptable to a relying party.

Examples of Web 2.0 SPs are Facebook and Google; both of these also act as IdPs for their domains, and potentially as IdPs for other domains, using OpenID. An example of an SP on an academic federation network, the UK Access Management Federation\footnote{http://www.ukfederation.org.uk/}, is the online publications database, ScienceDirect\footnote{http://www.sciencedirect.com/}; while the universities themselves are IdPs for their domains, and sometimes also SPs. In the first case, the `users' are internet users, in the second they are staff and students. In a digital ecosystem environment it is envisioned that all entities will have the ability to offer and consume services and act as IdPs for other entities, in social network of equals. Identity tasks in these various environments establish identity, assert attributes and handle authorisation.

\begin{figure*}
\begin{center}
\includegraphics[width=1.00\textwidth]{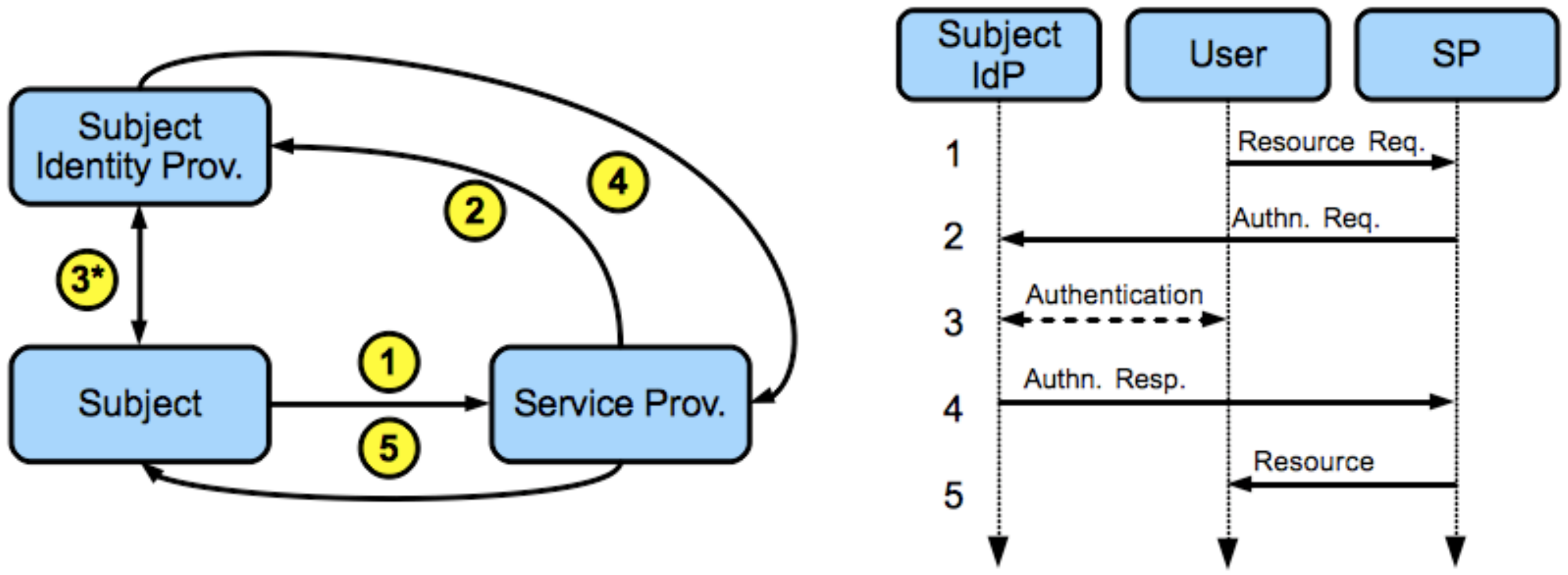}
\caption{High level view of common federation and user-centric \ac{SSO} protocols (e.g. SAML and OpenID).}
\label{fig:1}       
\end{center}
\end{figure*}

We begin our analysis by looking at trust requirements in federated and user-centric \ac{SSO}. \ac{SSO} is the most important identity task, and the killer application of these two identity technologies. In particular, we consider the SAML v2.0 \ac{SSO} profile and the OpenID v2.0 protocol. These two protocols, though differing in complexity (SAML is the more complex and extensible specification) and in terms of their transport layers (SAML uses various bindings, OpenID uses a HTTP GET Binding-like transport), they are sufficiently similar that we can examine them together\footnote{The transport protocol will not affect our analysis of the role of trust in identity tasks.}. We use `User', `Identity Provider' and `Service Provider'; where SAML uses `User-Agent',  `Identity Provider' and `Service Provider'; and where OpenID uses `User-Agent', `OpenID Provider' and `Relying Party', respectively. The basic protocol flow is given in Fig. \ref{fig:1}.

\begin{figure*}
\begin{center}
\includegraphics[width=0.55\textwidth]{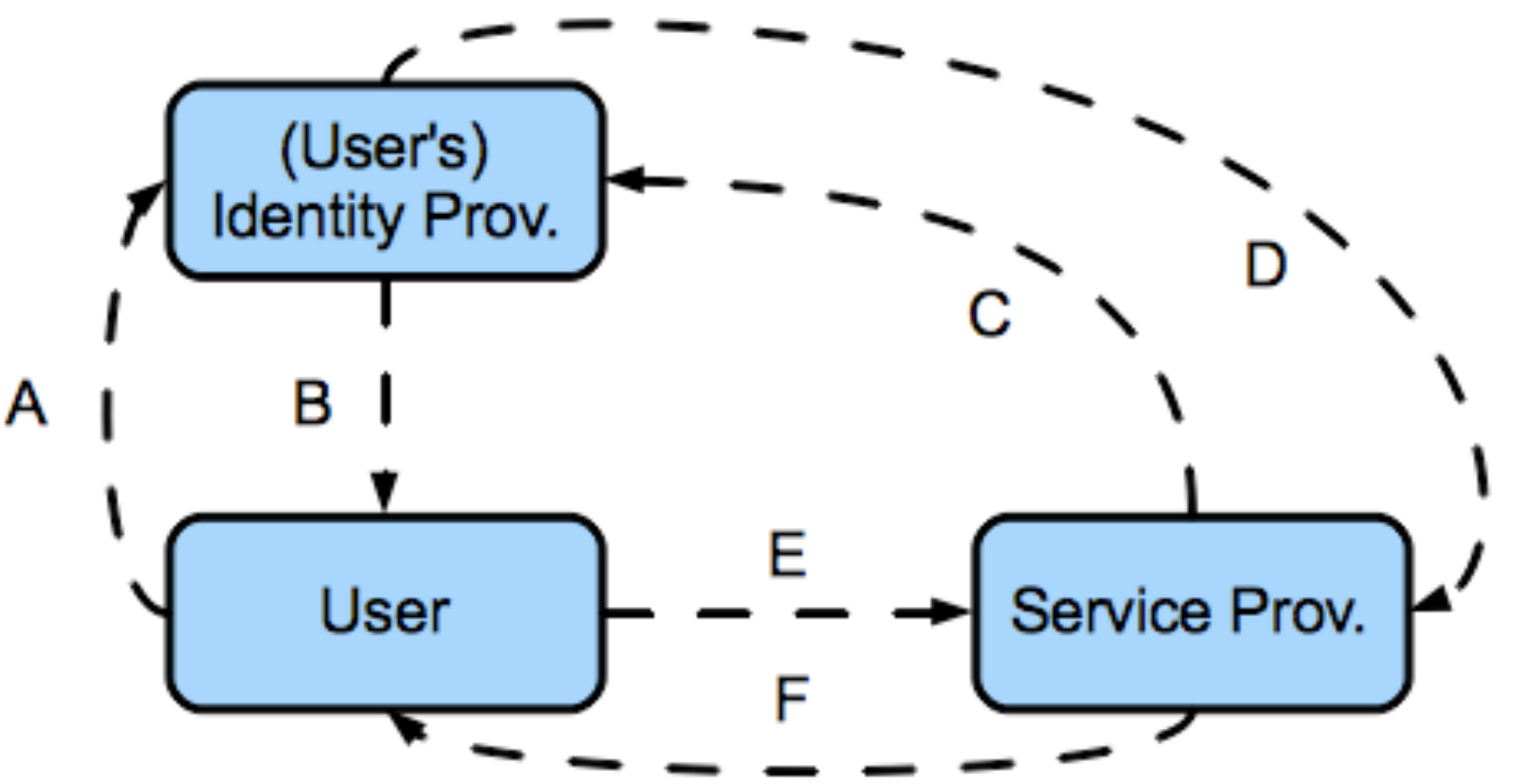}
\caption{Trust relationships in \ac{SSO} protocols.}
\label{fig:2}       
\end{center}
\end{figure*}

The trust requirements of identity management for protecting user privacy, when dealing with various IdPs and SPs in various identity scenarios are given in \citet{jsang_trust_2005}, whilst the trust requirements of IdP and SP in a range of similar scenarios are given in \citet{kylau_trust_2009}. Summarising and extrapolating from these sources, we look at the trust requirements of all parties involved in identity tasks, such as \ac{SSO}. Fig. \ref{fig:2} illustrates the trust relationships between the three parties in the \ac{SSO} identity task, which are explained below. These trust relationships are necessary for actors to interact with each other, because each of them are associated with a risk or set of risks. In each case, the context of the trust is given and the grounds for the trust. Where grounds are different in user-centric (UC) or federated (FD) identity, these are noted.
\begin{list}{\Alph{qcounter}:~}{\usecounter{qcounter}}
\item User trusts IdP to protect his privacy, to be secure and to manage his identity information appropriately*. (Context: identity provision. Grounds: agreement (FD) or user chooses IdP based on experience or reputation (UC).)
\item IdP trusts that user is who he claims to be, and that the user will manage granted credentials with care. (Context: identity self-assertion and responsibility. Grounds: authentication and agreement (FD) or positive risk assessment\footnote{In other words, the trustor (IdP) believes that the benefits of participation outweigh the risks, given that there are no strong assurances (as in the federated scenario).} (UC).)
\item SP trusts IdP to assert valid and truthful identity claims and to behave appropriately with data shared. (Context: make good assertions. Grounds: agreement and trust in methods\footnote{In other words, trust that secure technical measures and internal processes are in place.}.)
\item IdP trusts SP to adhere to privacy policies regarding the disclosure of user data. (Context: trust to maintain privacy. Grounds: agreement and trust in methods.)
\item User trusts SP not to correlate personal data about him from other IdPs\footnote{Resulting in two separate user identities becoming linked and the corresponding breach in privacy. (See \textit{unlinkability} \citep{pfitzmann_anonymity_2005}.)}*. (Context: trust to maintain privacy. Grounds: derived from agreement in D\dag (FD), trust in terms of service (UC).)
\item SP trusts user to abide by terms of service. (Context: trust in good intentions. Grounds: from agreement in C\dag (FD) or positive risk assessment (UC).)
\end{list}
*Additionally, users trust both IdP and SP to pass assertions only when requested by the user and that the mapping between user identities/data on both ends is correct. 
\dag In the federated scenario, trust relationships E and F are essentially derived from D and C, respectively, since federated agreements between IdP and SP and binding for all users in the IdP's identity domain.

These trust relationships may not be absolute, but they must be `sufficient', both individually and collectively, in order for the identity task to be possible. In general, trust in federated scenarios is based on formal agreements between the controllers of the identity domains that are being federated; trust is derived from the alliance. User-centric scenarios tend to rely on the less solid guarantees of reputation, in the case of users choosing an IdP, positive risk assessments in the case of IdPs and SPs trusting users, where risks of bad behaviour of unfaithful, anonymous users are in general outweighed by the benefits of a `leap of faith'. Larger risks require stronger trust relationships, and therefore stronger grounds for trust.

In DE's, and potentially on the Internet, entities can be different actors at different times. An entity can be a user, or subject, when establishing its identity, but may be an IdP or SP in other situations. It is reasonable, therefore, that all entities can either self assert their identities or have an IdP which can assert identities on their behalf. Conversely, it is reasonable that all entities can either consume identity assertions or have an IdP that can consume identity assertions on their behalf. In keeping with the philosophy of separating the roles of SP and IdP, we will advance the standard trust model of a federated/user-centric identity task as shown, by separating the SP into SP and SP's IdP, where the SP's IdP is capable of consuming assertions intended for the SP. For the remaining discussion we assume that IdPs are capable of both generating and consuming assertions on behalf of other entities. 

\begin{figure*}
\begin{center}
\includegraphics[width=0.60\textwidth]{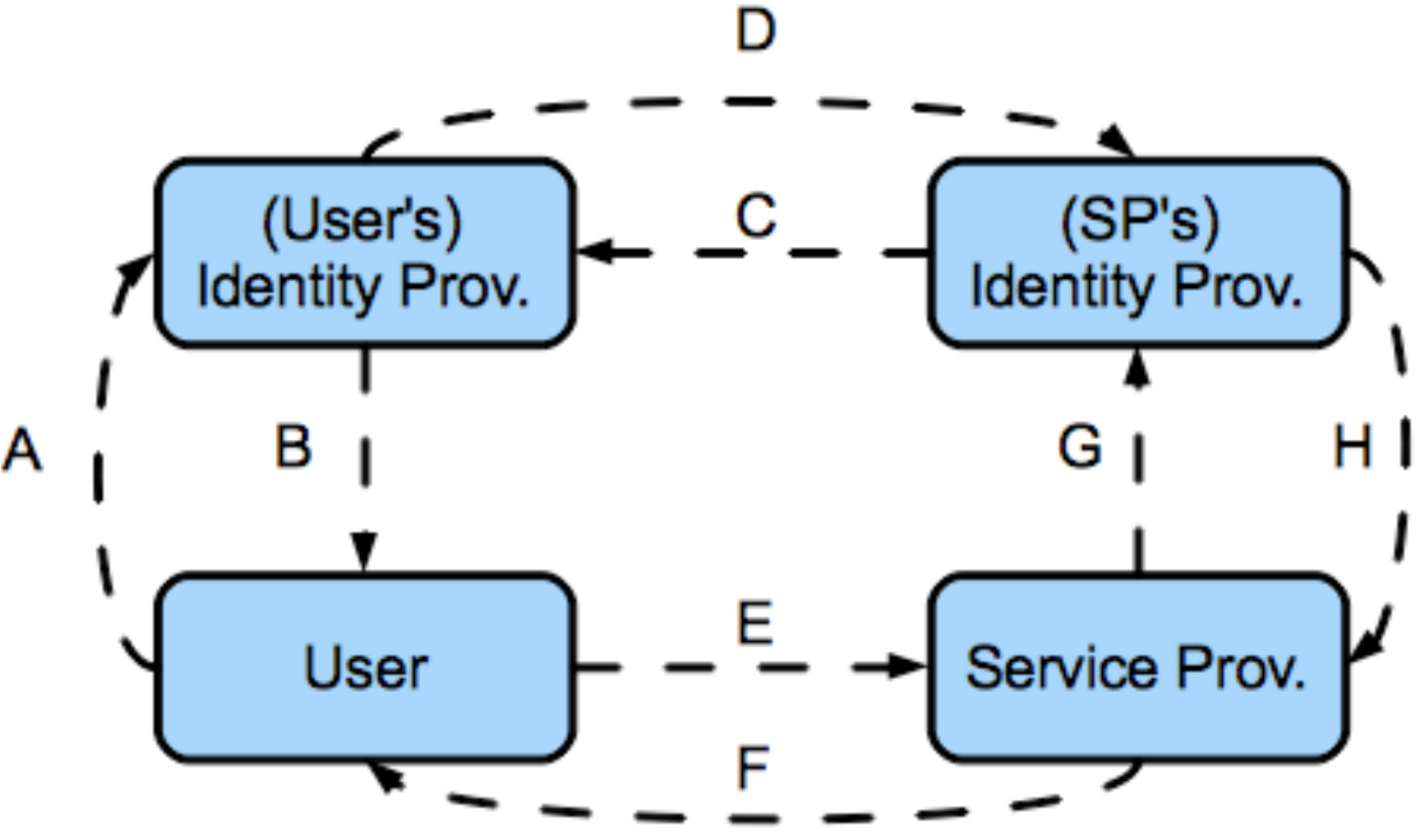}
\caption{Trust relationships in identity tasks in the general case.}
\label{fig:3}       
\end{center}
\end{figure*}
Fig. \ref{fig:3} illustrates the trust relationships of the general case as stated. The new trust relationships, G and H, are internal and absolute in the traditional federated/user-centric case, however, they may be between two different entities. If so, the analysis of these relationships will be similar to A and B. In the case where identity claims are self-asserted or self-verified, we assume that relationships A and B, or G and H, respectively, are internal and absolute. We will use this model for analysing the role of trust in decentralised identity in the proceeding discussion.

Trust relationships C and D, between the IdP and SP, are the most important, since these trust relationships are core to the federated agreement, in the federated scenario, or the primary leap of faith, in the user-centric scenario. The trust relationships between user and IdP are likely to pre-date, and to be at a higher level, than these trust relationships. Others SPs can be made known to the IdP, and vice versa, admitting new identity tasks, if trust relationships C and D are added for the new pair.
\section{Decentralised identity backed by trust}
\label{sec:3}
It is clear that the interplay between the various actors involved in an identity task is becoming more varied and complex over time, to address new business and collaboration scenarios, as reflected in the variety of federation topologies. \textit{IdentityFlow} is a software project designed to simplify the creation of complex, arbitrary, SAML-based protocols, to conduct arbitrary identity tasks. Each protocol must satisfy the trust requirements outlined in the general case in section \ref{sec:2} for the specific case of the identity task. We term these protocols, including the bindings,  the trust requirements of each actor and all logic for conducting the identity tasks, \textit{Identity Operations}. In this section, we outline how trust determines the success of identity operations and how trust transitivity is used to create a network of trust supporting operations.

\subsection{Trust in identity operations}

An agreement is explicit in federated scenarios, however, in place of an explicit agreement we can employ a mechanism for measuring trust. This allows for the following possibilities, i) that learning from experience can provide a basis for trust without having to formulate a fixed agreement, ii) that `trust ratings' of trust relationships might change based on dynamic factors and iii) that trust relationships might not exist directly between two actors, but might be derivable from a trust network based on referrals.

The evolution of trust ratings for relationships C and D lead to dynamic federations. It is necessary that C and D be rated sufficiently highly for a given operation to succeed. Sensitive operations, such as inter-organisation \ac{SSO}, will require high levels of trust, and perhaps an explicit agreement; less sensitive user-centric \ac{SSO} on online social networks, will not require such a high level of trust. 

The possibility of dynamic federations based of trust is also suggested by \citet{arias_cabarcos_enabling_2009}, who suggests extending Liberty's ``circle of trust'' model by encoding trust data in SAML assertions and using a ``trust engine'' to update the list of trusted identity domains held by IdPs/SPs. This list determines which identity domains can federate. This approach is quite practical and has a low impact on existing standards, however its notion of `trust' is not contextual, and therefore too coarse grained for the purposes of transitivity (see below). 
In our model, `trust checks' are conducted during protocol execution to ensure that trust relationships are sufficient. Each entity has has access to a \textit{trust manager} (see below). 
Calls to the trust manager during execution by trustee entity and context will yield a trust rating, which will either either exceed a trust threshold set for the trust relationship in the operation specification, or not.

\begin{figure*}
\begin{center}
\includegraphics[width=0.60\textwidth]{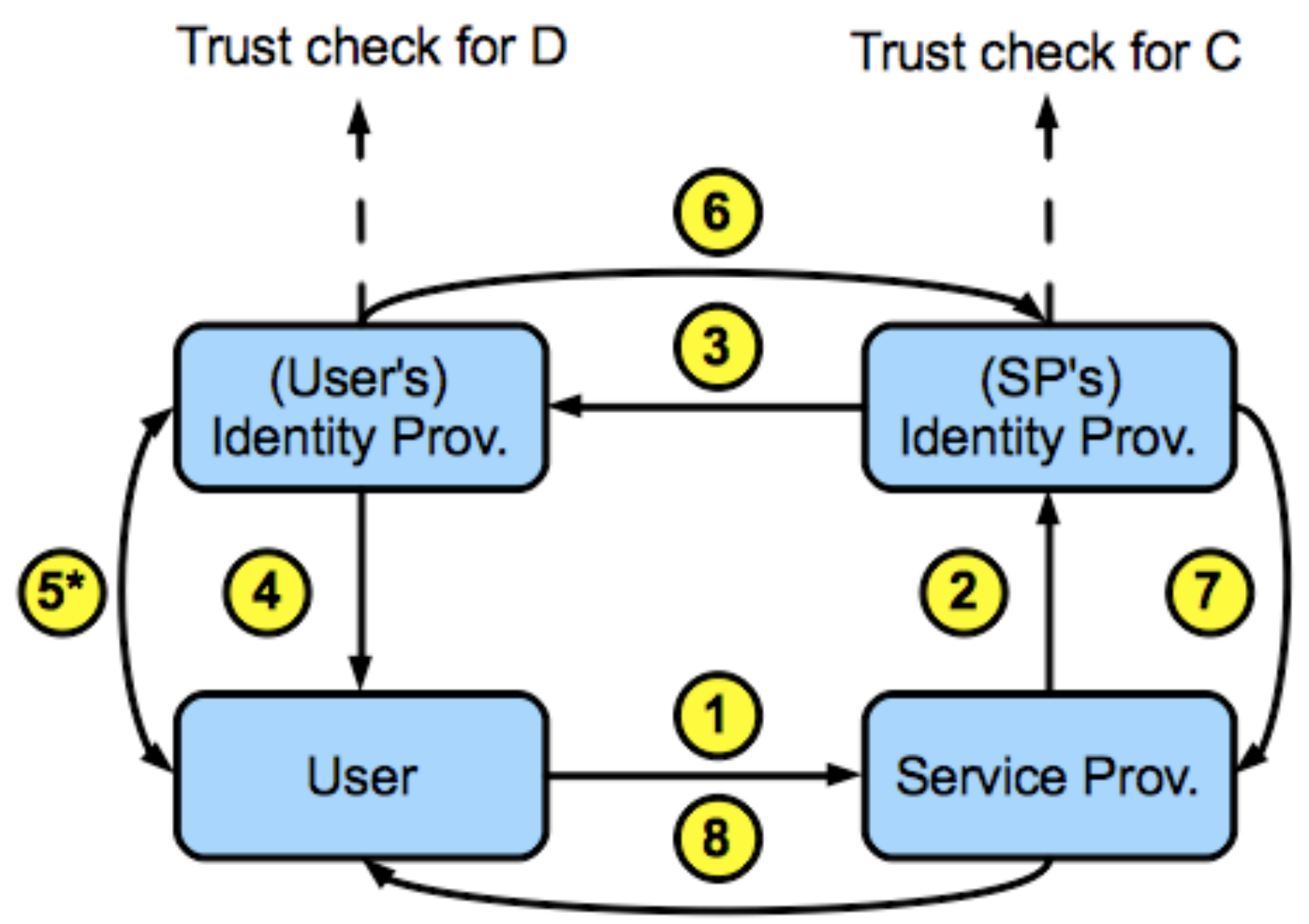}
\caption{Representative operation with trust checks during protocol execution.}
\label{fig:4}       
\end{center}
\end{figure*}

In order to analyse the protocol flow to identify points at which trust checks should be performed, we examine a generalised configuration of actors in an operation. The actors given in Fig. \ref{fig:3} in terms of their trust relationships is given again in Fig. \ref{fig:4} in terms of the sequence of connections between them in a generalised operation protocol flow. This protocol flow varies from the \ac{SSO} given in Fig. \ref{fig:1} only in that we allow for a physical separation of SP and SP IdP (which interprets assertions for the SP), therefore connections 2 and 7 merely reflect the passing of the authentication request from the SP to its IdP and the passing of the authentication response back from its IdP. This configuration of actors is essentially generic, as in the trust analysis, except that authentication requests and responses between the user's IdP and the SP's IdP may be passed through any combination of intermediaries\footnote{This should not present significant difficulties regarding security if messages are signed, encrypted and/or passed only to trusted parties, as appropriate.}.

During the execution of the protocol flow there will be appropriate points at which to measure the level of trust relationships to ensure they are sufficient for the operation to succeed, since messages from untrusted parties are useless. We have already identified trust relationships C and D (Fig. \ref{fig:3}) as crucial to federations and to the success of operations; now we examine the points at which their trust levels can be checked. There is no point in making connection 3 if the SP's IdP does have sufficient trust in the user's IdP in the context of `making good assertions', and similarly, there is no point in the user's IdP prompting the user to authenticate if it does not trust the SP's IdP in the context of `maintaing (user) privacy' respect the user's privacy. At these points, a failed authentication result should be passed back instead, and the operation should terminate. Trust checks can be used wherever it is felt that trust relationships are non-absolute and likely to evolve over time. We merely identified the most likely candidates as C and D. More complex protocol flows may perform additional or different trust checks between actors during protocol execution. 

\subsection{Trust networks from trust transitivity}
We begin our discussion of trust networks and trust transitivity by formally describing the role of the trust manager, which is a component responsible for providing evaluations of trust relationships, or trust ratings. Trust ratings are conveyed in (trustee,context) pairs, where trustee is the trustee and context is the trust context. The trust manager records `performance' trust ratings based on direct experience and is capable of gathering `referral' trust from third parties. Referrals are conveyed from an entity with performance trust in the trustee back to the trustor. Trust managers have the following functions,
\begin{enumerate}
\item\label{i1} Maintain a set of trust ratings with entities with whom the trustor has direct experience.
\item\label{i2} Discover trust transitive paths between trustor and trustee in the given context.
\item\label{i3} Aggregate these paths using appropriate strategies and algorithms to produce a trust rating.
\item\label{i4} Be capable of checking the integrity and authenticity of referrals from referees on the trust paths.
\item\label{i5} Have some mechanism for updating trust ratings based on experience.
\end{enumerate}

\ref{i2}. is a challenge in decentralised environments, or otherwise, where a virtual trust network must be traversed in order to discover trust transitive paths. A number of options for discovering such paths in decentralised networks with various topologies are given in \citet{e._ribeiro_de_mello_evaluation_2007}. A scheme for aggregating trust paths and using belief calculus to produce a compound rating is given in \citet{jsang_trust_2006}. \citet{jsang_semantic_2005} gives the rationale and methodology for verifying the integrity and authenticity of referrals.

\begin{figure*}
\begin{center}
\includegraphics[width=0.80\textwidth]{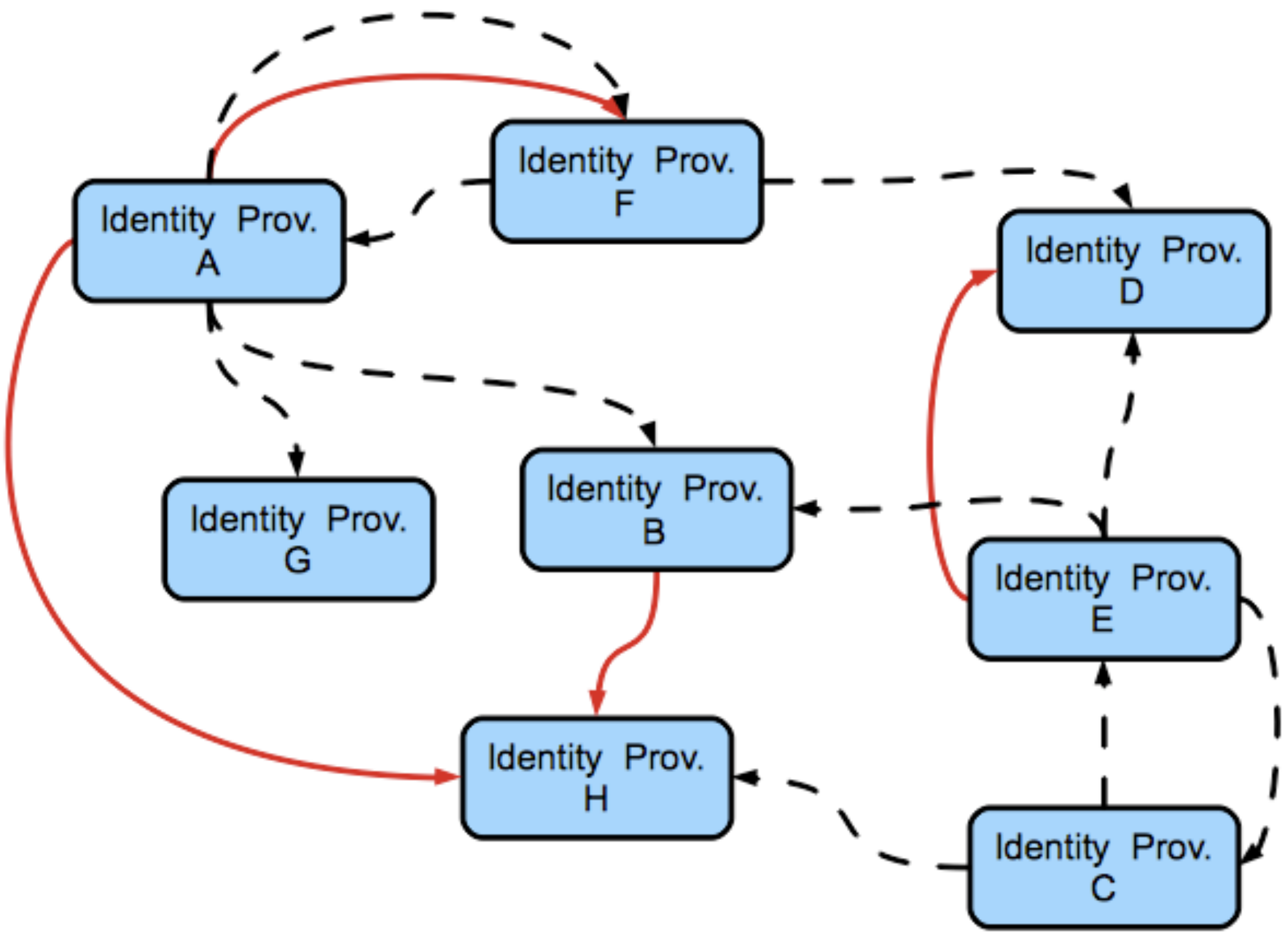}
\caption{Decentralised trust network of IdPs based on contextual trust relationships.}
\label{fig:5}       
\end{center}
\end{figure*}

Entities can update their trust ratings, and strategies for deriving trust ratings based on experiences from interacting with other entities. Where a referring entity is responsible for providing a trust rating, and where that trustee is found later to be undeserving of that rating, a suitable strategy may be deployed to reduce that referring entity's trust rating in referrals. `Experience reports' can be automatically or manually generated and submitted to the trust manager to derive an updated trust rating, according to some subjective scheme. 
The processes and algorithms for trust evaluation are described in \citet{mcgibney2007}. The open source project \textit{Trustflow}\footnote{TrustFlow, http://sourceforge.net/projects/trustflow/} is actively developing an implementation of a trust manager for use in P2P environments.

In this way, the trust manager constructs internal snapshots of portions of the trust network, allowing entities access to the trust ratings required to ensure that trust requirements in operations are satisfied. This trust network is constantly changing according to dynamic factors and subjective decisions.

By way of illustration, let us consider a portion of a decentralised network of IdPs, A-H, capable of acting as IdPs in a particular scenario. Performance trust relationships are represented by solid lines, whilst referral trust relationships are represented by dashed lines. We consider trust relationship C from Fig. \ref{fig:3} (although our analysis will apply equally to other relationships). C is the trust that the SP's IdP has in the user's IdP in the context of making good assertions. We are concerned, therefore, with the direct performance trust that the SP's IdP has in the user's IdP and, using trust transitivity, the chains of indirect referral trust from IdP to IdP, ending with a performance trust arc terminating at the user's IdP. We can infer a trust network from these relationships, illustrated in Fig. \ref{fig:5}.

\begin{figure*}
\centering
\mbox{\subfigure{\includegraphics[width=0.4\textwidth]{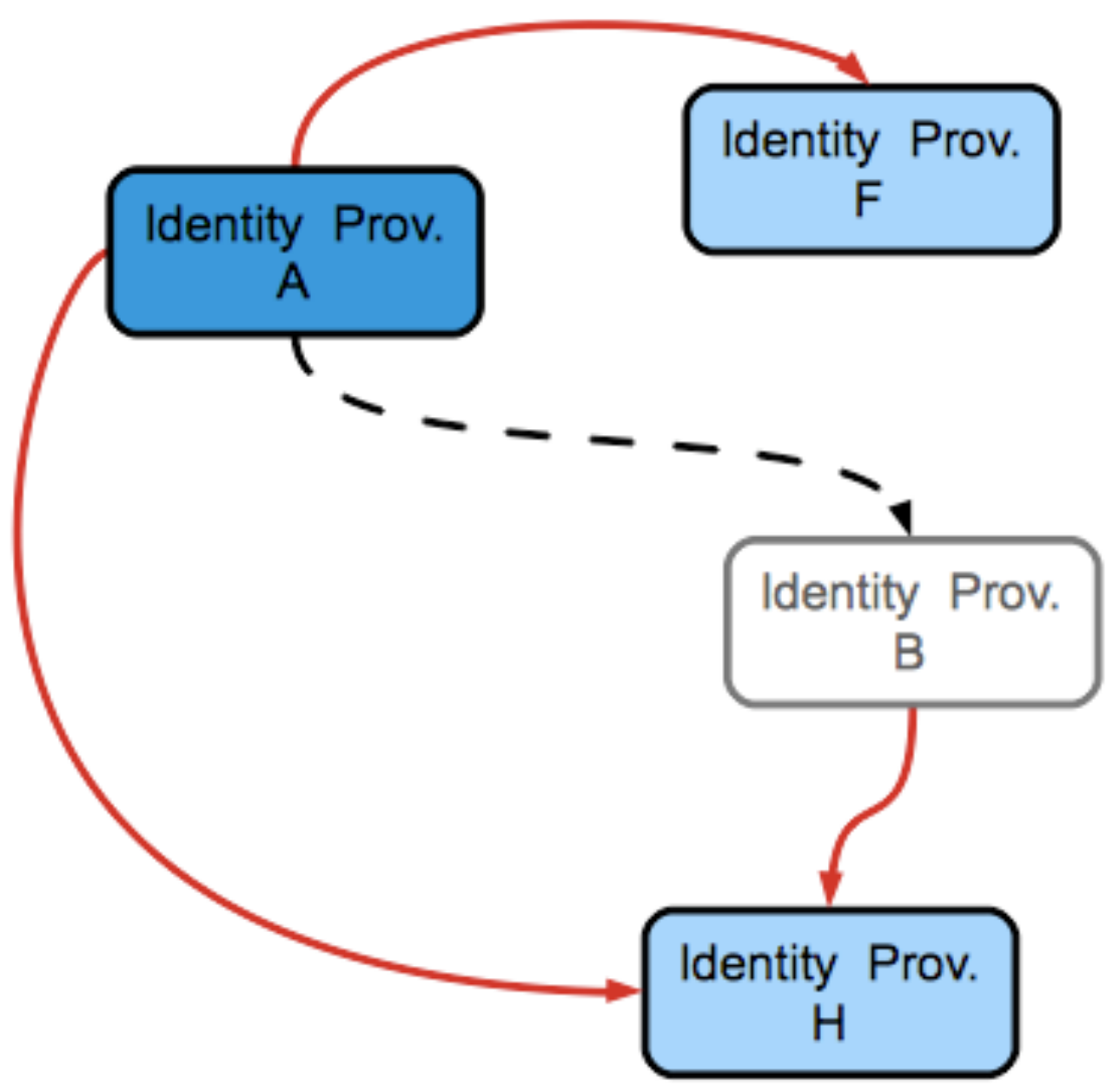}}\quad
\subfigure{\includegraphics[width=0.42\textwidth]{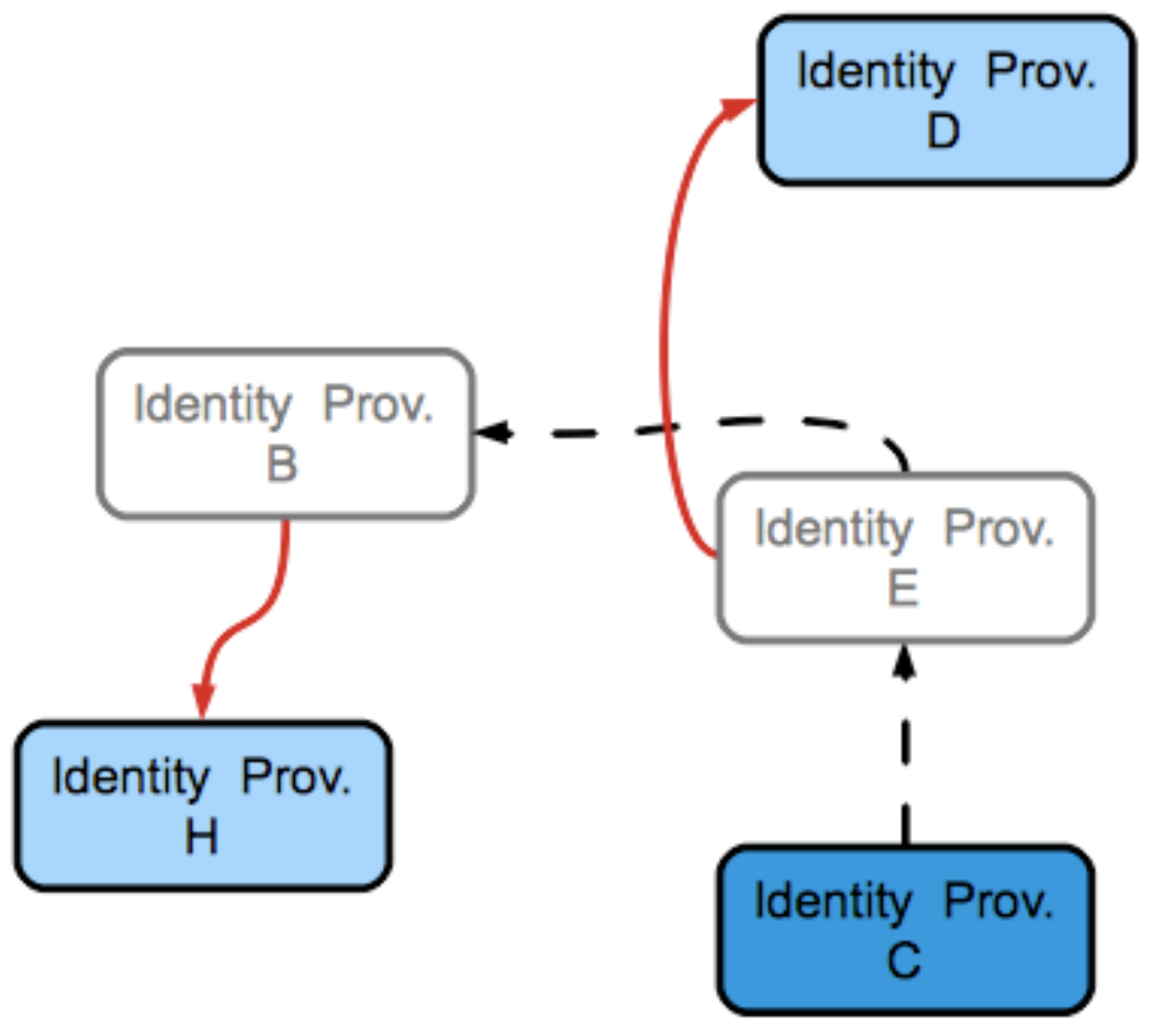} }}
\caption{Transitive, contextual trust graphs originating from IdPs A and C, respectively.} \label{fig:6}
\end{figure*}

Fig. \ref{fig:6} illustrates two portions of the trust network giving all possible trustees for trustors A and C respectively, by tracing valid trust transitive paths outwards from IdPs A and C. The lightly shaded boxes represent the trustees, while the transparent boxes are intermediate referees.
\section{Impact of decentralised identity on the Information Society}
\label{impact}

Fundamentally, we are considering the move from authority-centric centralised identity provisioning \citep{miyata_surveyidentity_2006} to user-centric \citep{maler_venn_2008} distributed identity provisioning \citep{markmcl_2009}, in which identity provision will be provided based on trust derived through social networks. 

\subsection{In the short to mid term}

The increasing availability of distributed identity provisioning will over time drastically change the \emph{landscape} of identity provisioning in the Information Society, supporting the ever-increasing trend \citep{pato2003identity} towards individuals and organisation using multiple identity provision schemes, including established centralised authorities, newer independent decentralised authorities, and future distributed trust-based self-provisioning of identity through social networks. The Web 2.0 phenomenon \citep{oreilly-web} has shown the potential and possibility for identity provisioning through trust in social networks \citep{andersen2007web}, i.e. trust networks. So, the future of the Information Society will involve extending Web 2.0 social networking to trust-based \citep{jsang_trust_2006} distributed identity provisioning.

\begin{figure}
\centering
\includegraphics[width=0.5\textwidth]{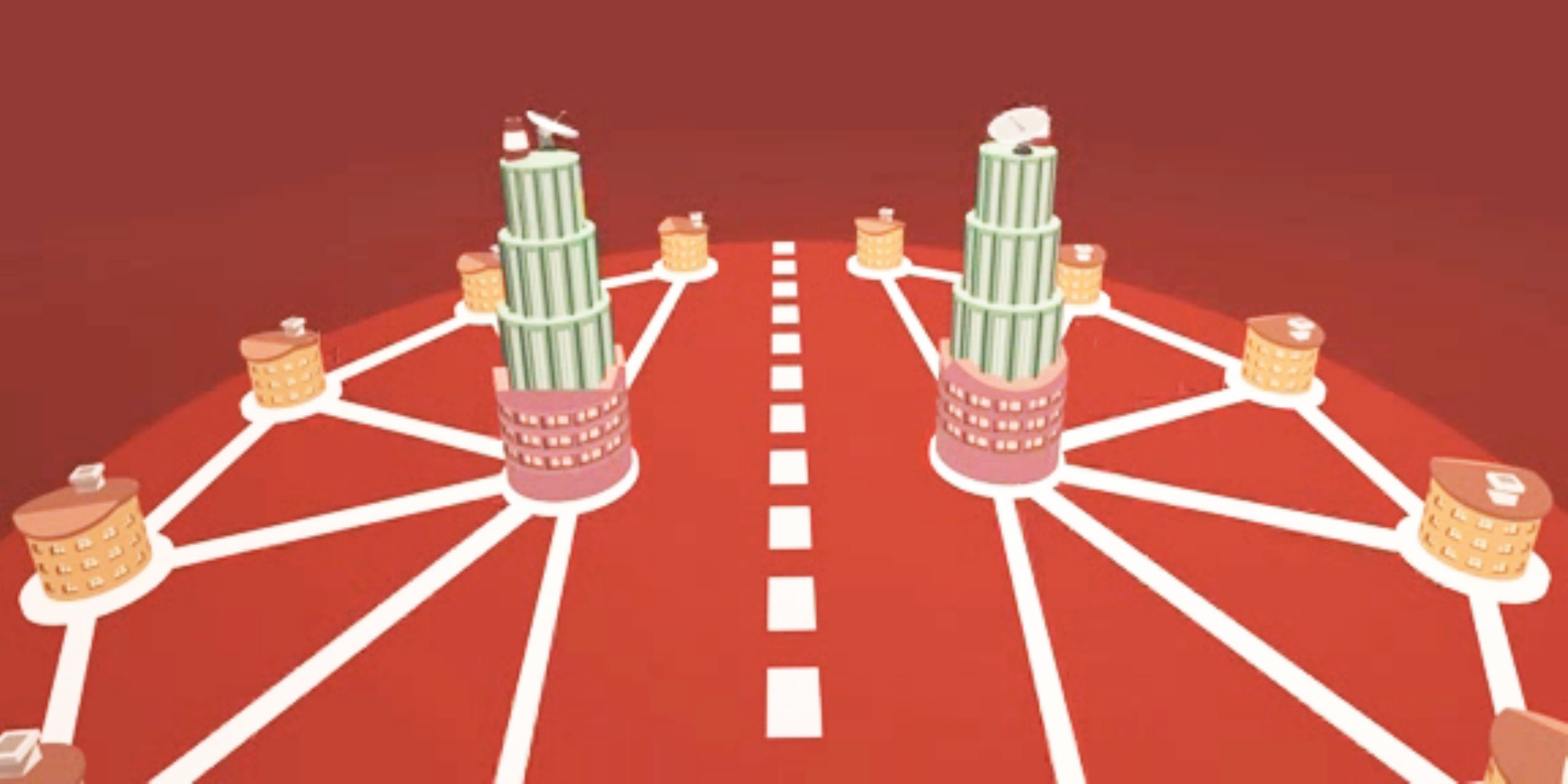}
\includegraphics[width=0.5\textwidth]{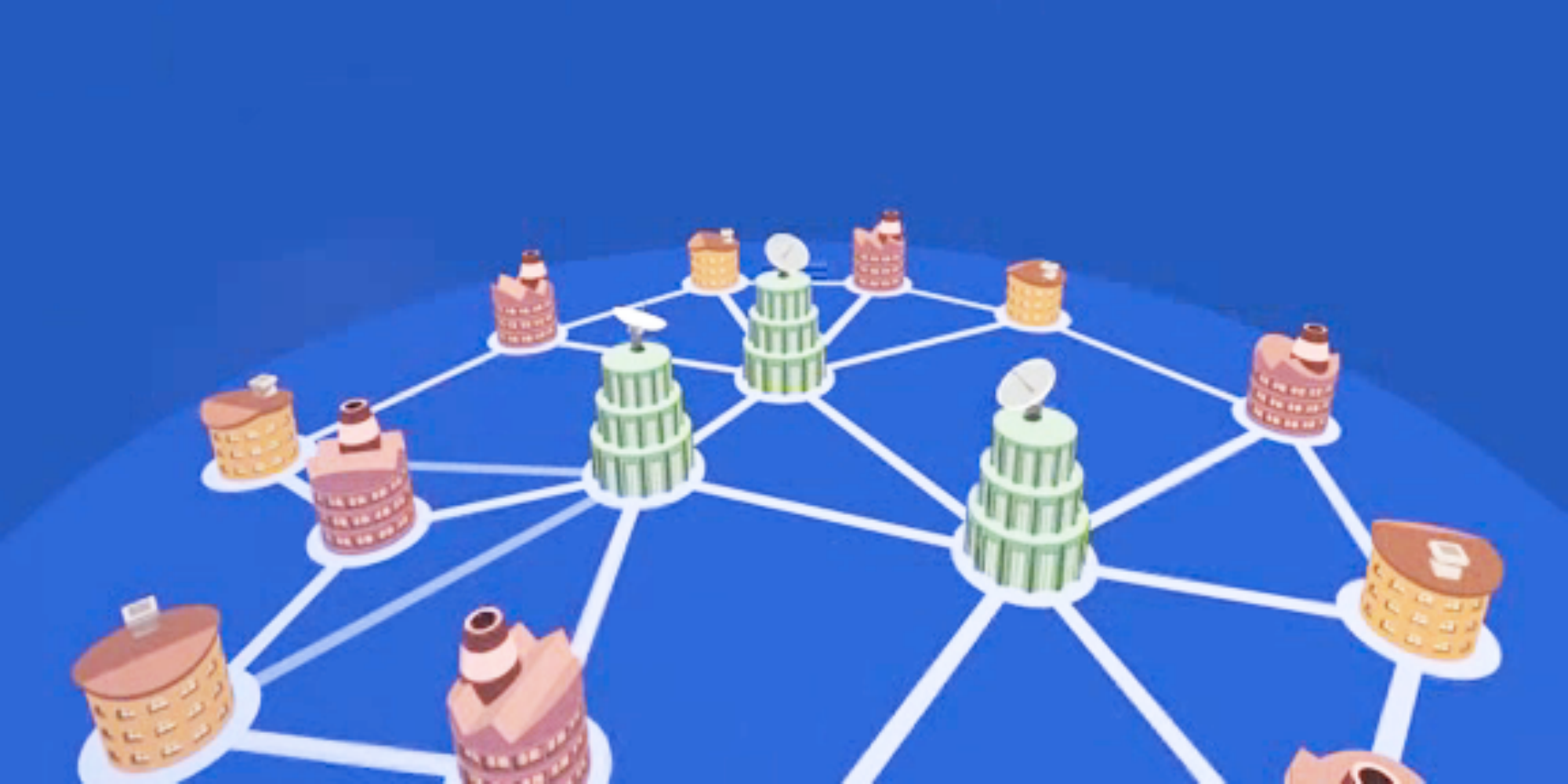}
\includegraphics[width=0.5\textwidth]{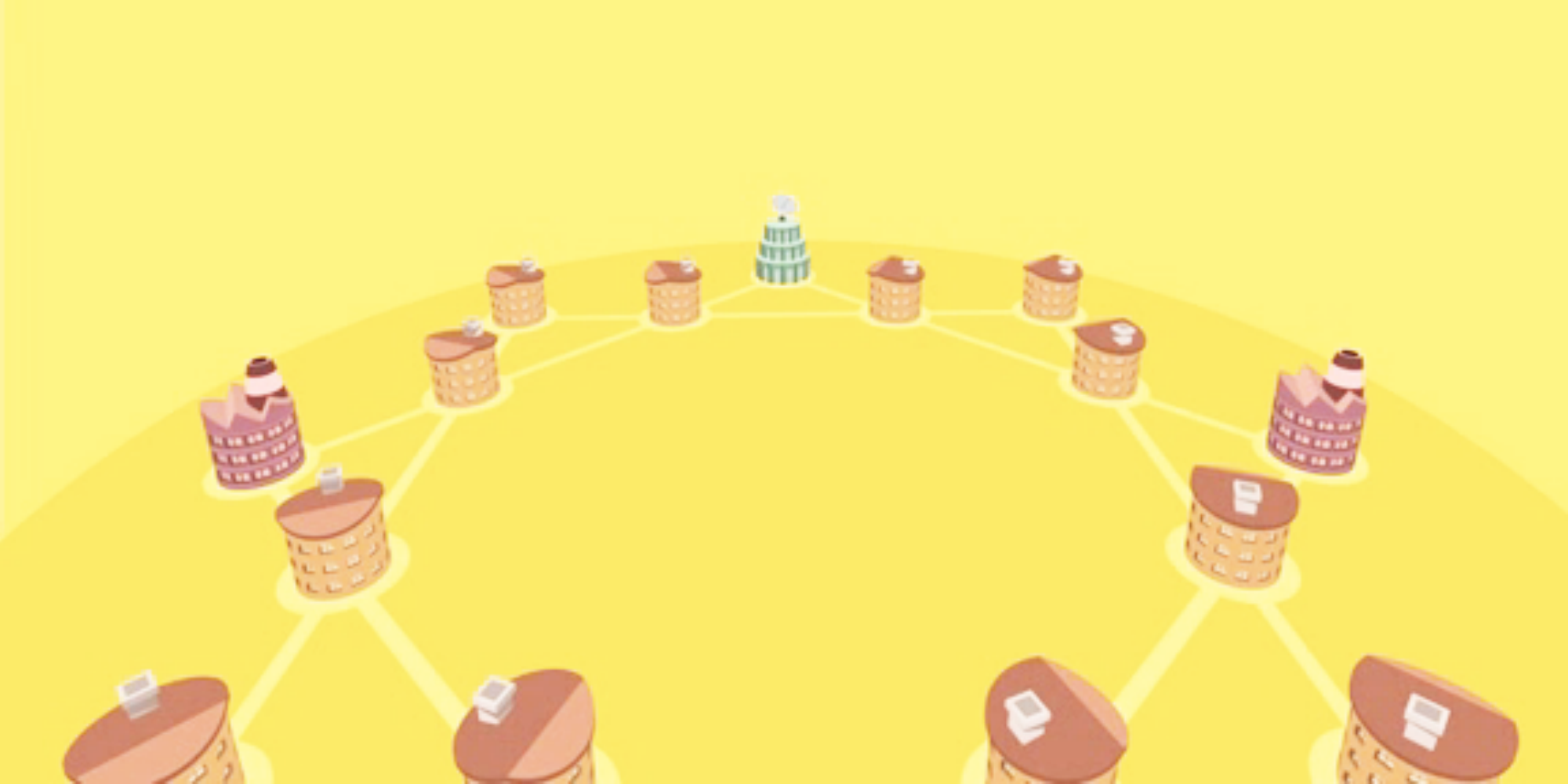}
\caption{Identity Provisioning Landscape (modified from \citep{wef}): The first figure provides a rendition of centralised identity provisioning as it has existed previously, which will continue being the dominant scheme in the immediate and short term \citep{miyata_surveyidentity_2006}. The second figure provides a rendition of decentralised identity provisioning \citep{sourceid_saml_2009}, which is an increasingly available and popular alternative in the short to mid term. Finally, the third figure provides a rendition of distributed identity provisioning \citep{markmcl_2009}, which will become increasingly available in the long term.}
\label{idenprovland}
\end{figure}

The availability of distributed identity provisioning will flatten the \emph{landscape} of identity provisioning, which was previously dominated and defined by centralised identity provisioning. So, past social and power structures \citep{pekrek_comparison_2009} of the \emph{landscape} will wain as individuals and organisations have an increasing choice of identity provisioning schemes \citep{johnson1001accountable,miyata_surveyidentity_2006}. This will allow a range of social and power structures for identity provisioning, appropriate for individuals, communities, and organisations. Ultimately, the balance of \emph{landscape} will shift, re-defined by the choice of individuals and organisations, from centralised identity provisioning to increasingly available decentralised identity provisioning (e.g. eduroam\footnote{Participating institutions, typically universities and other research and educational organisations, allows a user belonging to one institution to get network access when visiting another institution. The visiting user is authenticated in a decentralised manner by their home institution, and so using their existing identity \citep{florio2005eduroam}}.), and also distributed identity provisioning as it becomes available, as shown in Fig. \ref{idenprovland}.

\subsection{In the long term}

In this \emph{brave new world} each user would inherently possess a unique \emph{social identity} \citep{ashforth1989social,tajfel1974social} (distributed and relative in the context of social networks), which combined with distributed identity provisioning would lead to an inversion of the currently predominant membership model. So, instead of users registering for each website (or service) anew, they could simply add the website to their identity and grant access. Allowing users to have multiple services connected to their identity, instead of creating new identities for each service. This relationship is reminiscent of recent application platforms, such as Facebook's f8\footnote{http://www.facebook.com/f8} and Apple's App Store\footnote{http://www.apple.com/iphone/apps-for-iphone/}, but distributed in nature and so free from the control of centralised resource provisioning. Also, allowing for the reuse of the connections between users, akin to Google's Friend Connect \footnote{http://www.google.com/friendconnect/}, instead of reestablishing them for each new application. 

So, identity in the Information Society will arise naturally from the structure of the network, based on the relation of nodes to each other via social networks, or more generally \emph{networks of interaction} \citep{stevenson1990formal}, and therefore with the ability to scale and expand without centralised control, as shown in Fig. \ref{idenprovland}. Such distributed identity provisioning schemes will make use of the property that each node possesses a unique position in the network, i.e. sets of connections to other nodes in different networks. As such distributed identity provisioning will increasingly rely upon on \emph{challenge and response} style authentication mechanisms \citep{mitchell1989limitations}. A simple example would be additional authorisation requests (security checks) when connecting from relatively remote nodes (locations), i.e. from a different country, which shows that once node connectivity significantly changes the inherent identification it provides must be re-established. However, this will not preclude identity partitioning (audience segregation \citep{goffman59}) or multiple identities (e.g. work and personal), including specific identities for specialist communities, allowing for different kinds of social relationships to be established and maintained \citep{rachels75}.

\subsubsection{Digital Business Ecosystems}

We can further consider Digital Business Ecosystems (DBEs) \citep{dbebkintro} as a more concrete example of the potential impact of distributed identity provisioning on the Information Society. DBEs are distributed adaptive open socio-economic technical systems, with properties of self-organisation, scalability and sustainability, inspired by natural ecosystems. So, distributed identity provisioning would be a fundamental first step in creating dynamic virtual organisations (VOs) \citep{desanctis1999introduction} of \acp{SME} aiming to compete with established large \emph{keystone} firms \citep{iansiti2004kan}. Such VOs can makes use of distributed identity provisioning, in which all the members share identity provisioning tasks and activities, but may equally choose to adopt internal or external (non-competing) centralised identity provisioning (benevolent dictatorship approach \citep{johnson1001accountable}). However, proportional representation schemes \citep{johnson1001accountable} would be possible, in which some SMEs of a VO retain full control over their identity provisioning activities. Therefore, fundamentally, our research goes beyond just distributed identity provisioning, affording users choice over the social and power structures of their identity provisioning, rather than having centralised identity provision thrust upon them.

\begin{figure*}
\centering
\includegraphics[width=1.00\textwidth]{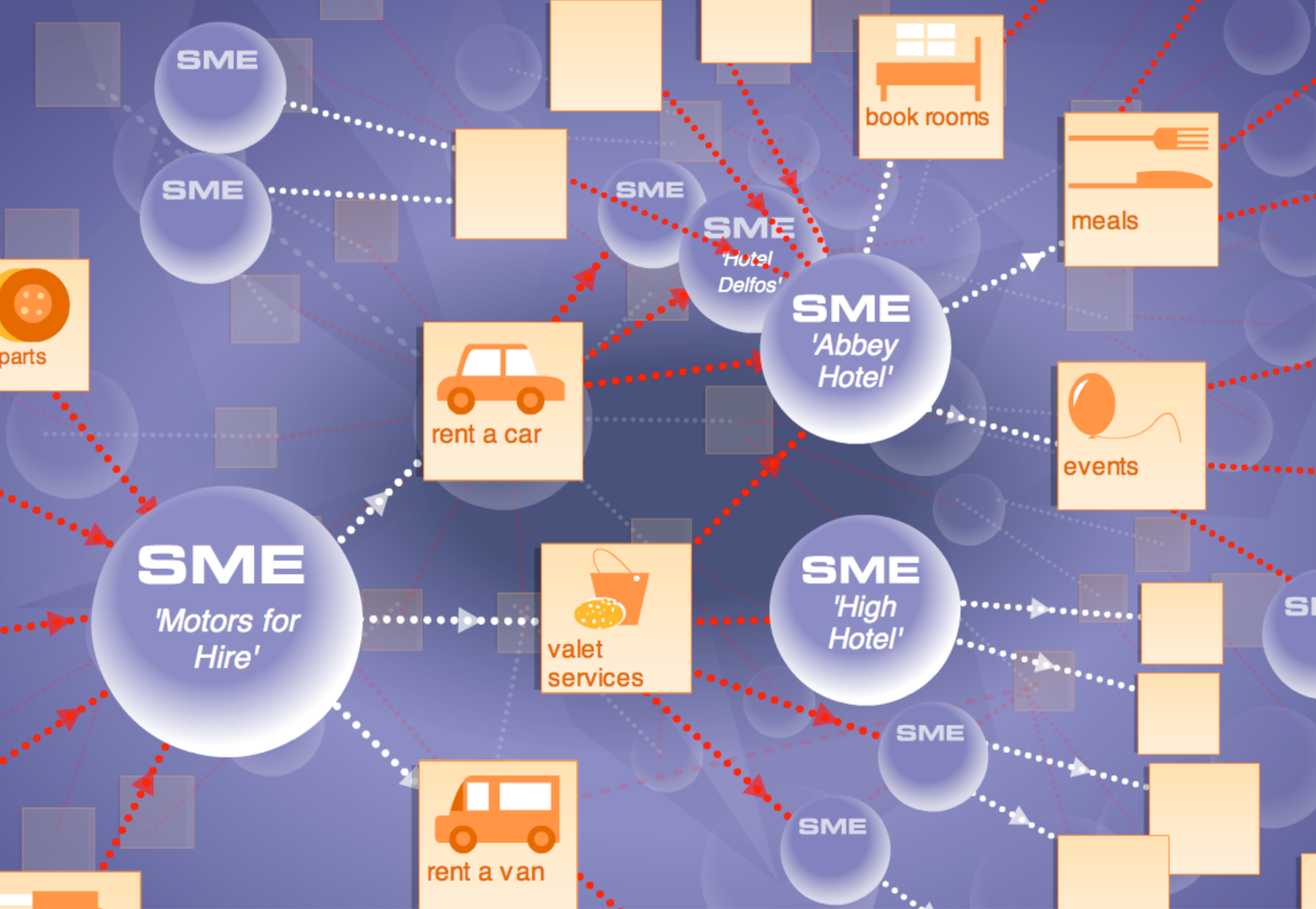}
\caption{Digital Business Ecosystem \citep{dbebkintro}: Conceptual visualisation \citep{dbevis} showing a DBE of interacting \acl{SME} users, via the services they provide and consume. Creating networks of business ecosystems distributed over different geographical regions, business domains, and industry sectors.}
\label{DBEprojectFinal}       
\end{figure*}
\section{Conclusions and future work}
\label{conclusions}
We have seen how a network of trust can be used to give trust ratings between a trustor and a trustee in identity contexts, such as `making good assertions' or `recommending good IdP', how a trust manager can maintain and gather trust ratings, and how trust transitivity allows us to trace and aggregate trust paths in the trust network. We have also seen that operations, such as \ac{SSO}, rely on a set of implicit trust relationships, some of which form the foundation of federations, dynamic or otherwise. By extension it is clear that federations can be formed between the identity domains of any two IdPs, given sufficient trust. Furthermore, we have discussed how this decentralised identity can arise, both in general terms, based on existing trends of ever-increasing diversity in the provisioning of identity, and specific examples such as \emph{business ecosystems} of SME networks.

Future work includes the integration of the open source \textit{IdentityFlow} and \textit{TrustFlow} projects to enable third parties to realise trustworthy, decentralised identity management. There is also considerable scope for future work in dynamic identity provisioning in unstable coalitions. Furthermore, the development of the trust manager to draw trust inferences across context boundaries and to infer trust from social networks presents a number of interesting possibilities.

\section{Acknowledgments}
This work is funded by the EU FP6 Network of Excellence OPAALS, http://www.opaals.org

\bibliographystyle{spbasic}
\bibliography{sigproc,references}

\end{document}